\documentstyle[12pt]{article}

\newcommand{\bigzerou}{%
\smash{\lower1.7ex\hbox{\bg 0}}}
\renewcommand{\theequation}{\arabic{section}.\arabic{equation}}

\newcommand{\half}{\frac{1}{2}}

\newcommand{\nonum}{\nonumber}

\newcommand{\mapright}[1]{%
\smash{\mathop{%
\hbox to 1.0cm{\rightarrowfill}}\limits^{#1}}}
\newcommand{\mapleft}[1]{%
\smash{\mathop{%
\hbox to 1.3cm{\leftarrowfill}}\limits^{#1}}}

\newcommand{\no}{\nonumber}

\begin{document}


\begin{titlepage}
\vglue 3cm

\begin{center}
\vglue 0.5cm
{\Large\bf $N=4$ Supersymmetric Yang-Mills Theory
\\ on Orbifold-$T^4/{\bf Z}_2$} 
\vglue 1cm
{\large Masao Jinzenji${}^\dagger $,  Toru Sasaki${}^* $} 
\vglue 0.5cm
{\it ${}^\dagger $
Division of Mathematics, Graduate School of Science,
                    Hokkaido University
,Sapporo 060-0810 Japan 
}\\
{\it ${}^*$Department of Physics,
Hokkaido University,Sapporo 060-0810,Japan}
{\it ${}^\dagger $jin@math.sci.hokudai.ac.jp \\
${}^*$ sasaki@particle.sci.hokudai.ac.jp}

\baselineskip=12pt

\vglue 1cm
\begin{abstract}
  We derive the partition function of $N=4 $
supersymmetric Yang-Mills theory
on orbifold-$T^4/{\bf Z}_2 $.
In classical geometry, 
$K3$ surface is constructed from the orbifold-$T^4/{\bf Z}_2 $.
Along the same way as the orbifold construction, 
we construct the partition function of $K3$ surface 
from orbifold-$T^4/{\bf Z}_2$.
The partition function is given by the product of
the contribution of the untwisted sector of $T^4/{\bf Z}_2$,
and that of the twisted sector of $T^4/{\bf Z}_2 $
i.e.,  ${\cal O}(-2) $ curve blow-up formula.
\end{abstract}
\end{center}
\end{titlepage}
\section{Introduction}
\label{sec:1}
Since the work of Vafa and Witten \cite{vafa-witten} have appeared, twisted 
$N=4$
super Yang-Mills theory becomes a laboratory of testing $S$-duality 
conjecture, and it is now one of the concrete examples of the duality 
conjectures in various string theories. 
In \cite{vafa-witten} two major conjectures are made:\\
(i) The partition function of
twisted $N=4$ theory on 4-manifold is
described as the summation of the Euler number of the ASD
instanton moduli space with gauge group $G$.\\
(ii) Under $S $-duality transformation
this partition function is transformed to
the partition function with its dual gauge group ${\hat G} $.
This duality is originated from Montonen-Olive duality \cite{M-O}.\\
For (i), it is confirmed that twisted $N=4$ theory is described as
a balanced topological field theory and that 
(i) is surely true if certain vanishing theorem holds 
\cite{laba}, \cite{dijkgraaf}.
In the spirit of (i), matter coupled version is also considered
and several analysis are done in \cite{sako}.

Vafa and Witten derived the partition function 
satisfying both (i) and (ii) by physical method.
(This partition function is called Vafa-Witten formula.)
In their derivation,
the most interesting part is the part which divides the evaluation of 
partition functions into two sectors: the bulk part and the cosmic string 
part. This observation arises from the mass perturbation of the theory.
This perturbation is realized by a section of the canonical bundle of the 
base 4-manifold (complex surface) $X$, and there appears drastic difference 
between the zero-locus of $K_{X}$  
and the other part of $X$ (they call these locus ``cosmic string'' and 
``bulk'' respectively). Note that this observation was also applied to 
the topological Yang-Mills theory by Witten \cite{witten}.
Next, they seem to identify the contribution 
from the cosmic string with the (generalized) blow-up formula 
speculated from the mathematical result of 
Yoshioka, and the one from the bulk with the Hilbert scheme of the points 
on the virtual manifold $X_{0}$ that has  Euler number $3(\chi(X)+\sigma(X))$ 
and trivial canonical bundle. The principle of adding up these contributions
is given by the celebrated $S$-duality conjecture. 

With these results, it seems that the partition function of twisted 
$N=4$ super Yang-Mills theory on any 4-manifold (at least algebraic)
can be computed by combining Hilbert scheme of points of a certain surface
with some blow-up formula. Then we try to interpret the orbifold construction
of Kummer surface ($K3$ surface) from Abelian surface $T^{4}$ as the 
combination of these two contributions. In this construction, we first divide 
$T^{4}$ by ${\bf Z}_{2}$. This process gives us the manifold $S_{0}$ 
with Euler number 
$8$ with trivial canonical bundle and sixteen orbifold singularities.
After that,  we blow-up the sixteen orbifold singularities and 
obtain $K3$ surface. Then we speculate that we can realize these processes 
at the level of the partition function of twisted $N=4$ SYM theory 
like the discussion given in \cite{vafa-witten}. But there are slight 
differences between our speculation and the original discussion in 
\cite{vafa-witten}. In the orbifold process, the blow-up operation should 
not change the canonical bundle mainly because $T^{4}$ and $K3$ have both 
trivial canonical bundles. So we must change the form of blow-up formula 
slightly in this case. Mathematically speaking, it is the difference 
between ${\cal O}(-1)$ curve blow-up and ${\cal O}(-2)$ curve blow-up. 
With these blow- up formulas, we can reconstruct the partition 
function of $K3$ surface from the 
partition function of $S_{0}$, which is fundamentally equivalent to the 
G\"ottsche formula of $S_{0}$. Coincidence with the well-known formula 
of $K3$ partition function in \cite{vafa-witten} is derived from the 
mathematical equalities among eta and theta functions. 
We think that our derivation is one example of orbifold construction of 
$4$-manifold at the level of partition function of $N=4$ SYM, and we expect 
many applications.

This paper is organized as follows. 

In Section 2, we review the classical geometry of the orbifold construction 
of $K3$ from $T^{4}/{\bf Z}_{2}$.

In section 3. we introduce G\"ottsche formula of the singular 4-fold 
$S_{0}=T^{4}/{\bf Z}_{2}$ and construct the partition function of $S_{0}$
(we call this part untwisted sector).

In section 4, we propose the form of ${\cal O}(-2)$ curve blow-up formula 
(we call this part twisted sector) and derive the partition function 
of $K3$ by successive application of the ${\cal O}(-2)$ curve blow-up formula 
to the partition function of $S_{0}$.

\section{Classical Theory of $T^4/{\bf Z}_2$}
\label{sec:2}

In this section, we review the 
geometry of the manifold  
$T^4/{\bf Z}_2$ \cite{Fukaya}, \cite{Asp}.

First, we construct $K3$ surface
from $T^4 $ \cite{Asp}.
It is well known that compact complex K\"ahler 4-manifold 
with trivial canonical bundle $K_X=0 $ is only $K3$ or $T^4$ \cite{Fukaya}. 
$
T^4=C^2/{\bf Z}^4
$
is realized as a quotient space  of the complex plane 
$
(z_1,z_2)\in C^2
$
divided by ${\bf Z}^4 $ action,
where 
${\bf Z}^4$ action is $z_k\to z_k+1,z_k\to z_k+i,k=1,2 $.
To obtain $K3$ surface, we first divide $T^4 $
with ${\bf Z}_2 $ action, where
${\bf Z}_2$ action is given by $(z_1,z_2)\to (-z_1,-z_2)$
,(we denote $T^4/{\bf Z}_2=S_0$ in the following).
Then one gets 
sixteen fixed points of ${\bf Z}_2 $ action 
\begin{equation}
(0,0),\;(0,\frac{1}{2}),\;(0,\frac{i}{2}),\;
\ldots,\;(\frac{1}{2}+\frac{i}{2},\frac{1}{2}+\frac{i}{2}).
\label{fix}
\end{equation}
Let us consider the topology of $T^4/{\bf Z}_2$. 
Since the ${\bf Z}_2$ action conserves the complex structure
and leaves the K\"ahler form invariant,
we expect $S_0$ to be a complex  K\"ahler orbifold.
Also, a moment's thought shows that any 
non-contractable loops of $T^4 $
shrink to points after the ${\bf Z}_2$
identification.
Thus $\pi_1 (S_0)=0 $. 
Moreover, the holomorphic 2-form $dz_1\wedge dz_2 $ is invariant.
Thus we obtain $K_{S_{0}}=0 $.

Since $S_{0}$ has sixteen $Z_{2}$ orbifold 
singularities coming from the fixed points 
in (\ref{fix}), 
we have to resolve these singularities.
As we explain below, we can resolve the orbifold singularities
by ${\cal O}(-2) $ curve blow-up so that
$K_X(=0) $ is kept \cite{Asp}.
In general, blow-up is done by
replacing a point with a curve. 
In this case, we replace an orbifold singularity 
with a ${\cal O}(-2) $ curve $E$.
Let us first consider the blow-up of one ${\bf Z}_{2}$ singularity 
$\pi: \tilde{X}\rightarrow X$.
Then we assume, 
\begin{equation}
K_{\tilde X}=\pi^{*}K_X+n\cdot E. 
\label{k1}
\end{equation}
Using the adjunction formula, we have 
\begin{equation}
2g_{E}-2=(E+K_{\tilde{X}})\cdot E.
\label{k2}
\end{equation}
Since $g_{E}=0 $ and $E$ is ${\cal O}(-2) $ curve,
then we obtain,
\begin{equation}
K_{\tilde X}\cdot E=(\pi^{*}K_X+nE)\cdot E=-2n=0.
\label{k3}
\end{equation}
Thus we can conclude 
\begin{equation}
K_{\tilde X}=\pi^{*}K_X(=0).
\label{k4}
\end{equation}
Hence ${\cal O}(-2)$ curve blow-up does not  change the canonical bundle
$K_{X}$. 
Using this ${\cal O}(-2) $ curve blow-up,
we can resolve the sixteen orbifold singularities,
and we get the smooth 4-manifold with 
trivial canonical bundle. This surface is nothing but the $K3$ surface. 

These processes are equivalently replaced with 
the following processes \cite{Fukaya}.
First we remove 16 4-ball $B_{4}$'s from $T^4$, which are neighborhood 
of 16 fixed points 
and divide $(T^4-16 B_{4})$ with ${\bf Z}_2$ action.
Then the boundaries of $(T^4-16 B_{4})/{\bf Z}_{2}$  
turn into sixteen $RP^{3}$'s.
On the other hand, we have Ricci flat space $T^{*}S^{2}$, that also has 
$RP^{3}$ as the boundary. Therefore, we can glue $(T^4-16 B_{4})/{\bf Z}_{2}$ 
with sixteen $T^{*}S^{2}$'s along sixteen $RP^{3}$'s smoothly.
In this way, we obtain $K3$ surface, which is also a Ricci flat complex
surface. 

Next, we discuss the 
cohomology of $T^4/{\bf Z}_2 $.
Especially we repeat the previous discussion with Hodge diamond.

First we write down the Hodge diamond of $T^4$.
\begin{equation}
\left.
  \begin{array}{ccccc}
 & &h^{0,0}& & \\
 &h^{1,0}& &h^{0,1}&\\
h^{2,0}& &h^{1,1}& &h^{0,2}\\
 &h^{2,1}& &h^{1,2}& \\
 & &h^{2,2}& & \\
  \end{array}
\right.
=
\left.
  \begin{array}{ccccc}
 & &1& &\\
 &2& &2&\\
1& &4& &1\\
 &2& &2&\\
& &1& & \\
  \end{array}
\right.
\end{equation}
The Hodge diamond of $S_0$ (modulo torsion)
is obtained from the one of $T^{4}$ by picking up ${\bf Z}_{2}$ 
invariant forms 
of $T^{4}$. We call this Hodge diamond untwisted sector.
\begin{equation}
\left.
  \begin{array}{ccccc}
 & &h^{0,0}& & \\
 &h^{1,0}& &h^{0,1}&\\
h^{2,0}& &h^{1,1}& &h^{0,2}\\
 &h^{2,1}& &h^{1,2}& \\
 & &h^{2,2}& & \\
  \end{array}
\right.
=
\left.
  \begin{array}{ccccc}
 & &1& &\\
 &0& &0&\\
1& &4& &1\\
 &0& &0&\\
& &1& & \\
  \end{array}
\right.
\end{equation}

After sixteen blow-ups, we obtain additional sixteen $(1,1)$ forms, and 
the Hodge diamond changes into,  

\begin{equation}
\left.
  \begin{array}{ccccc}
 & &h^{0,0}& & \\
 &h^{1,0}& &h^{0,1}&\\
h^{2,0}& &h^{1,1}& &h^{0,2}\\
 &h^{2,1}& &h^{1,2}& \\
 & &h^{2,2}& & \\
  \end{array}
\right.
=
\left.
  \begin{array}{ccccc}
 & &1& &\\
 &0& &0&\\
1& &20& &1\\
 &0& &0&\\
& &1& &. \\
  \end{array}
\right.
\end{equation}
Note that this Hodge diamond is 
the same as that of $K3$.

\section{Untwisted Sector}
\label{sec:3}
\setcounter{equation}{0}

In this section, we derive the partition function
of the untwisted sector of $T^4/{\bf Z}_2 $ \cite{vafa-witten}, \cite{mukai}, 
\cite{yoshiell}, \cite{qin}, \cite{nak}.

\subsection{G\"ottsche Formula}
\label{sec:3-1}

In this subsection, we introduce G\"ottsche formula of $X$
\cite{mukai},\cite{qin}.

Following \cite{qin},
we first introduce virtual Hodge number $e^{s,t}(X) $
and virtual Hodge polynomial $e(X;x,y) $ of $X$.
\\
The virtual Hodge number is labeled by a pair of integers $(s,t)$ and is 
given by
\begin{equation}
e^{s,t}(X)=\sum_k(-1)^kh^{s,t}(H^k_c(X,Q)),
\label{ho}
\end{equation}
and
the virtual Hodge polynomial of $X$ is given by
\begin{equation}
e(X;x,y)=\sum_{s,t=0}^2e^{s,t}(X)x^sy^t.
\label{po}
\end{equation}
Note that the virtual Hodge polynomial can be viewed
as a convenient tool for computing the Hodge numbers
of smooth projective varieties. 
Main reason of such characteristics comes from the following well-behavior:\\
(i) When $X$ is projective and smooth, 
$e(X;x,y)$ is the usual Hodge polynomial.\\
(ii) If $Y$ is a Zariski-closed subscheme of $X$, then 
\begin{equation}
e(X;x,y)=e(Y;x,y)+e(X-Y;x,y).
\end{equation} 
(iii) If $f:X\rightarrow Y $ is a Zariski-locally trivial bundle with fiber 
$F$, then 
\begin{equation}
e(X;x,y)=e(Y;x,y)\cdot e(F;x,y).
\end{equation} 
For any algebraic surface $X$, G\"ottsche formula describes the generating 
function of the virtual Hodge polynomial of Hilbert scheme $X^{[n]}$:
\begin{eqnarray}
Z^X(\tau;x,y)
&\equiv&\sum_{n=0}^\infty q^n e(X^{[n]};x,y)
\nonum
\\
&=&
\prod_{m=1}^\infty  \prod_{s,t=0}^2
(1-x^{s+m-1}y^{t+m-1}q^m)^{(-1)^{s+t+1}h^{s,t}(X)},
\end{eqnarray}
where $h^{s,t}(X) $ stands for the Hodge numbers of $X$.

When we apply the formula to 
$T^4 $, we get
\begin{eqnarray}
& &
Z^{T^4}(\tau;x,x)
\nonum
\\
&=&\sum_{n=0}^\infty q^n e((T^4)^{[n]};x,x)
\nonum
\\
&=&
\prod_{m=1}^\infty  
\frac
{
(1-x^{2m-1}q^m)^4
(1-x^{2m+1}q^m)^4
}
{
(1-x^{2m-2}q^m)
(1-x^{2m}q^m)^6
(1-x^{2m+2}q^m)
}.
\end{eqnarray}
In the $K3$ case, 
we have,
\begin{eqnarray}
& &
Z^{K3}(\tau;x,x)
\nonum
\\
&=&\sum_{n=0}^\infty q^n e((K3)^{[n]};x,x)
\nonum
\\
&=&
\prod_{m=1}^\infty  
\frac
{
1
}
{
(1-x^{2m-2}q^m)
(1-x^{2m}q^m)^{22}
(1-x^{2m+2}q^m)
}.
\end{eqnarray}
Now, we apply  
G\"ottsche formula to compute the contribution from 
the untwisted sector of
$T^4/{\bf Z}_2 $. 
\begin{eqnarray}
& &
Z^{S_0}(\tau;x,x)
\nonum
\\
&=&
Z^{T^4}(\tau;x,x)|_{{\bf Z}_{2}~{\rm inv.}}
\nonum
\\
&=&
\left.
\prod_{m=1}^\infty  
\frac
{
(1-x^{2m-1}q^m)^4
(1-x^{2m+1}q^m)^4
}
{
(1-x^{2m-2}q^m)
(1-x^{2m}q^m)^6
(1-x^{2m+2}q^m)
}\right|_{{\bf Z}_{2}~{\rm inv.}}
\nonum
\\
&=&
\prod_{m=1}^\infty  
\frac
{
1
}
{
(1-x^{2m-2}q^m)
(1-x^{2m}q^m)^6
(1-x^{2m+2}q^m)
}
\end{eqnarray}
Following \cite{vafa-witten},
we define 
\begin{eqnarray}
G(\tau)&\equiv& q^{-\frac{1}{3}}Z^{S_0}(\tau;x=y=1)
\nonum
\\
&=&
\frac{1}{\eta^8(\tau)},
\label{8}
\end{eqnarray}
for later use.
In the next subsection, we derive the partition function 
of the untwisted sector, which is fundamentally constructed from the formula
(\ref{8}). 

\subsection{Partition Function of the Untwisted Sector of $S_0 $}
\label{sec:3-2}

In this subsection, we derive the partition function of
$S_0$ of the untwisted sector. 
\cite{qin}, \cite{vafa-witten}, \cite{yoshiell}, \cite{yoshi2}.

\paragraph{General Structure of Vafa-Witten Conjecture}
~\\
Following \cite{qin}, we review the
general structure of Vafa-Witten conjecture.

For $SO(3)$ theory with second Stiefel-Whitney class $v$ on $X$,
partition function is defined by, 
\begin{equation}
Z^X_v(\tau)\equiv q^{-{\frac{\chi(X)}{12}}}\sum_k \chi({\cal N}(v,k))q^k,
\end{equation}
where ${\cal N}(v,k)$ is 
the moduli space of anti-self-dual connections
associated to $SO(3)$-principal bundle with
second Stiefel-Whitney class $v\in H^2(X,Z_2)$ 
and fractional instanton number $k\in Z/4$,
and $\chi(X)$ is Euler number of $X$.
For this partition function, Vafa and Witten conjectured
\begin{equation}
Z^X_v\left(-\frac{1}{\tau}\right)=2^{-\frac{b_2(X)}{2}}
\left(
\frac{\tau}{i}
\right)^{-\frac{\chi(X)}{2}}
\cdot
\sum_{u\in H^2(X,Z_2)}
(-1)^{u\cdot v}Z^X_u(\tau).
\label{vw}
\end{equation}
For later use, we introduce
\[
Z_{SU(2)}^X(\tau)\equiv \half Z_0^X(\tau),
\]
\begin{equation}
Z_{SO(3)}^X(\tau)\equiv \sum_{u\in H^2(X,Z_2)} Z_u^X(\tau).
\end{equation}
For these partition functions, the conjecture is reduced to the following 
formula:
\begin{equation}
Z^X_0\left(-\frac{1}{\tau}\right)=2^{-\frac{b_2(X)}{2}}
\left(
\frac{\tau}{i}
\right)^{-\frac{\chi(X)}{2}}
Z_{SO(3)}^X(\tau).
\label{so3}
\end{equation}

\paragraph{$SO(3)$ Bundles on Spin Manifold $X$}
~\\
In this part, we discuss
the typical properties of $SO(3) $ bundles on
the spin manifold $X$. 
Following \cite{vafa-witten}, we first point out
\begin{equation}
k=n-\frac{v^2}{4},
\end{equation}
where $n$ is second Chern class 
associated with rank two vector bundle,
and
$k$ takes value on $k\in {\bf Z}/2 $
according to the  relation $v^2=0,\;2\; {\rm mod}~4$
on a spin manifold.
As $H^2(X,Z) $ is $b_2(X)$ dimensional
(and torsion free),
$v$ can take $2^{b_2(X)}$ values.
In $SO(3) $ theory, we must sum over them.

There is no need to study separately $2^{b_2(X)}$
because $X $ 
has a very large 
diffeomorphism group which permutes the possible values of $v$.
One obvious diffeomorphism invariant of  $v$ is 
the value of $v^2$ modulo 4;
if it is $0$ we call even and if it is 2
we call odd.
If $v$ is odd, it is certainly non-zero,
but for $v$ even there is one more invariant:
whether $v$ is zero of not.
It turns out that up to diffeomorphsm,
the invariants just stated are the only invariants of $v$.
So on $X$ 
there are really three partition functions
to compute, namely the partition functions 
for $v=0$ even but non-zero, and odd.
We call these $Z^X_0,Z^X_{even},$ and $Z^X_{odd}$. 
Similarly, we write $n^X_0,n^X_{even},$ and $n^X_{odd}$
for the number of values of $v$
that are, respectively, trivial, even but non-trivial, and odd.
The number of each type is counted in Appendix A for $K3$ case.
One can think of three types of partition functions:
\begin{eqnarray}
&&Z^X_0(\tau)=q^{-\frac{\chi(X)}{12}}\sum_n\chi({\cal N}(0,n))q^n,
\nonumber\\ 
&&Z^X_{even}(\tau)=q^{-\frac{\chi(X)}{12}}\sum_n\chi({\cal N}(v_{even},n))q^n,
\nonumber\\ 
&&Z^X_{odd}(\tau)=q^{-\frac{\chi(X)}{12}}
\sum_n\chi({\cal N}(v_{even},n))q^{n-\half}.
\end{eqnarray} 

\paragraph{Partition Function of the Untwisted Sector of $S_0$}
~\\
In this part,
we derive the partition function of the untwisted sector of $S_0$
as a concrete example.
First we think of the moduli space of 
$X=T^4 $ 
as an example of Corollary 1.7 in \cite{yoshiell}.
In \cite{yoshiell},
they treat the moduli space ${\cal M}_H(c_1,c_2) $
of rank two stable sheaves $E$
with chern classes $c_1,c_2 $,
but we want to treat the moduli space ${\cal N}(v,k) $
of $SO(3) $ or $SU(2) $ vector bundles with $v,k $.
Thus following \cite{qin},
we identify ${\cal M}_H(c_1,c_2)\equiv {\cal N}(v,k) $
with $c_1=v~{\rm mod}~2 $ and $k=c_2-\frac{v^2}{4} $
in the following. 
According to \cite{yoshi2},
the moduli space of rank two stable sheaves $E$ of $V$
is given below. We introduce Mukai vector 
\begin{equation}
V=ch(E)=2+c_1+\frac{c_1^2-2c_2}{2},
\end{equation}
and its inner product
\begin{eqnarray}
<V^2>&=&-\int_X
\left(2+c_1+\frac{c_1^2-2c_2}{2}\right)\vee
\left(2+c_1+\frac{c_1^2-2c_2}{2}\right)
\nonum
\\
&=&4c_2-c_1^2.
\end{eqnarray}
Note that we use a symmetric bilinear form on 
$\oplus{}_i H^{2i}(X,Z) $:
\begin{eqnarray}
<x,y>&=&-\int_X(x\vee y)
\nonum
\\
&=&
  \int_X(x_1y_1-x_0y_2-x_2y_0),
\end{eqnarray}
where $x=x_0+x_1+x_2,x_1\in H^{2i}(X,Z) $
and $x\vee =x_0-x_1+x_2$.

We get the moduli space labeled by $V$
\begin{equation}
M_H^X(V)\cong {\hat X}\times (X)^{[\frac{<V^2>}{2}]}
= {\hat X}\times(X)^{[2c_2-\frac{c_1^2}{2}]},
\end{equation}
where $H$ is some lime bundle on $X$ and
${\hat X}$ is the dual of $X$ 
\cite{yoshi2}.
\\
Since $S_{0}$ has the trivial canonical bundle like $K3$ and $T^{4}$, it 
doesn't have cosmic strings, which are given by zero locus of the section of 
the canonical bundle.
Hence we assume the moduli space of $V$
in $S_0 $ case,
\begin{equation}
M_H(V)\cong (S_0)^{[\frac{<V^2>}{2}]}
= (S_0)^{[2n-\frac{v^2}{2}-1]},
\end{equation}      
where we use the vanishing Picard group
to drop the ${\hat X} $ part
and  above identifications.
Note that the shift in $[\cdots  ] $
by $-1$
comes from $\chi({\cal O}_{S_0})=
\frac{1}{12}\int_{S_{0}}((c_{1}(S_{0}))^{2}+c_{2}(S_{0}))=\frac{2}{3} $
instead of $\chi({\cal O}_{T^4})=0 $
(see also the definition of the moduli space in \cite{yoshiell}).
Using this we define the partition function of
$S_0$ of the untwisted sector 
with  even type,
\begin{eqnarray}
Z^{S_{0}}_{even}(\tau)
&=&
q^{-\frac{2}{3}}\sum_{v^2\equiv 0(mod~4),n} e(M_H(V))q^n  
\nonum
\\
&=&
q^{-\frac{2}{3}}\sum_n e((S_0)^{[2n-1]})
q^n
\nonum  
\\
&=&
q^{-\frac{1}{6}}\sum_m e((S_0)^{[m]})
\frac{(q^{\frac{1}{2}})^m-(-q^{\frac{1}{2}})^m}{2}
\nonum
\\
&=&
\frac{1}{2}G(\frac{\tau}{2})
-\frac{1}{2}e^{\frac{\pi i}{3}}G(\frac{\tau}{2}+\frac{1}{2}).
\label{zse}
\end{eqnarray}

Similarly we define the partition function of
$S_0$ of the untwisted sector 
for odd type,
\begin{eqnarray}
Z^{S_{0}}_{odd}(\tau)
&=&
q^{-\frac{2}{3}}\sum_{v^2\equiv 2(mod~4),n}e(M_H(V))q^{n-\frac{1}{2}}  
\nonum
\\
&=&
q^{-\frac{2}{3}}\sum_n
e((S_0)^{[2n-2]})
q^{n-\frac{1}{2}}  
\nonum
\\
&=&
q^{-\frac{1}{6}}\sum_m e((S_0)^{[m]})
\frac{(q^{\frac{1}{2}})^m+(-q^{\frac{1}{2}})^m}{2}
\nonum
\\
&=&
\frac{1}{2}G(\frac{\tau}{2})
+\frac{1}{2}e^{\frac{\pi i}{3}}G(\frac{\tau}{2}+\frac{1}{2}).
\label{zso}
\end{eqnarray}
For 0 type, we follow \cite{vafa-witten} and set,
\begin{equation}
Z^{S_{0}}_0(\tau)=NG(2\tau)+\frac{1}{2}G(\frac{\tau}{2})
-\frac{1}{2}e^{\frac{\pi i}{3}}G(\frac{\tau}{2}+\frac{1}{2}),
\end{equation}
where $N$ is not determined. Note that $G(2\tau)$ is obtained from  
$G(\frac{\tau}{2})$ using the modular transformation 
$\tau \rightarrow -\frac{1}{\tau}$.

Strictly speaking, $S_0$ is not a spin manifold 
since the intersection matrix of $S_{0}$ is given by 
\begin{equation}
\left(\begin{array}{cc}0&\half\\
                       \half&0\\ \end{array}\right)
\oplus 
\left(\begin{array}{cc}0&\half\\
                       \half&0\\ \end{array}\right)
\oplus 
\left(\begin{array}{cc}0&\half\\
                       \half&0\\ \end{array}\right),
\end{equation}
\cite{priv}.
Hence it is possible for  $v^2$ to take values $1$ and $3$ mod $4$.
This tells us that the general story in the previous part
is not applicable directly.
However, for our purpose to investigate the connection
between $K3$ and $S_0$,
it is sufficient to forget the contribution from $v^2 =1$ or $3 $ mod $4$
(detailed explanation is done in \ref{sec:4-2}).

In the rest of this part,
we discuss the modular properties 
of these partition functions
for later use.

For $\tau\to \tau+1 ,$
\begin{eqnarray*}
  Z^{S_{0}}_{even}(\tau+1)
&=&
\frac{1}{2}G(\frac{\tau}{2}+\frac{1}{2})
-\frac{1}{2}e^{\frac{\pi i }{3}}
e^{-\frac{\pi i}{12}8}G(\frac{\tau}{2})
\nonum
\\
&=&
e^{-\frac{4\pi i }{3}}\left(
\frac{1}{2}G(\frac{\tau}{2})
-\frac{1}{2}e^{\frac{\pi i }{3}}G(\frac{\tau}{2}+\frac{1}{2})
\right),
\end{eqnarray*}
\begin{eqnarray}
  Z^{S_{0}}_{odd}(\tau+1)
&=&
\frac{1}{2}G(\frac{\tau}{2}+\frac{1}{2})
+\frac{1}{2}e^{\frac{\pi i }{3}}
e^{-\frac{\pi i}{12}8}G(\frac{\tau}{2})
\nonum
\\
&=&
-e^{-\frac{4\pi i }{3}}\left(
\frac{1}{2}G(\frac{\tau}{2})
+\frac{1}{2}e^{\frac{\pi i }{3}}G(\frac{\tau}{2}+\frac{1}{2})
\right).
\end{eqnarray}

For $\tau$ to $-\frac{1}{\tau},$
\begin{equation}
Z^{S_{0}}_{even}\left(-\frac{1}{\tau}\right)
-
Z^{S_{0}}_{odd}\left(-\frac{1}{\tau}\right)
=
-e^{\frac{\pi i}{3}}G(-\frac{1}{2\tau}+\half)
=
\left(
\frac{\tau}{i}
\right)^{-4}
\left(
Z^{S_{0}}_{even}(\tau)
-
Z^{S_{0}}_{odd}(\tau)
\right)
\label{zsm}
\end{equation}

\section{Twisted Sector}
\label{sec:4}
\setcounter{equation}{0}
In this section, we introduce ${\cal O}(-2)$ blow-up formula and reconstruct 
$K3$ partition function \cite{yoshioka}, \cite{qin}, \cite{vafa-witten}.
\subsection{${\cal O}(-2) $ Curve Blow-up }
\label{sec:4-1}
In this subsection, we make an ansatz for the ${\cal O}(-2)$ blow-up formula 
for the partition function. We assume that the blow-up formula in this case 
is given by the following three functions:
\begin{equation}
\frac{\theta_{2}(\tau)}{\eta(\tau)^{2}},\;\; 
\frac{\theta_{3}(\tau)}{\eta(\tau)^{2}},\;\;
\frac{\theta_{4}(\tau)}{\eta(\tau)^{2}}.   
\end{equation}
In \cite{vafa-witten}, the blow-up formula for ${\cal O}(-1)$ curve 
is used to describe the effect of the existence of the cosmic string, 
that is the zero-locus of a section of the canonical divisor.  
This formula is proved by Li and Qin \cite{qin}.
Of course, their results are inspired by the work of Yoshioka \cite{yoshioka}. 
In the ${\cal O}(-1)$ curve case, the formula is given by,
\begin{equation} 
\frac{\theta_{2}(2\tau)}{\eta(\tau)^{2}},\;\;\;  
\frac{\theta_{3}(2\tau)}{\eta(\tau)^{2}}.
\label{-1}
\end{equation}
 Modification of 
$\theta(2\tau)$ into  $\theta(\tau)$ can be estimated by following the proof 
of the ${\cal O}(-1)$ curve blow-up formula in \cite{qin}, changing the 
condition of $E\cdot E = -1$  into $E\cdot E = -2$. But we don't have the 
explicit proof of this ansatz. We leave it to the future work.
The reason for introducing $\theta_{4}(\tau)$ 
comes from the fact that 
$\theta_{3}(\tau)$ has both integral and half integral powers of $q$.
We need $\theta_{4}(\tau)$ to project out half-integral (resp. integral)
powers of $q$.

The factor $\frac{1}{\eta(\tau)^{2}}$ in (\ref{-1}) comes from the fact that 
we use  
Gieseker-Maruyama compactification of the moduli space. It appears as the 
effect of the additional cohomology class $E$ and of the fact that 
vector bundle we treat has two components. Hence it does not depend on the 
self intersection of the class $E$, and we do not change this factor.    
 
With these ansatz, we define $\theta$ functions
\begin{equation}
\Theta_\pm(\tau)\equiv 
\half\left(
\theta_3^8(\tau)\pm \theta_4^8(\tau)
\right),
\end{equation}
\begin{equation}
\Theta_2(\tau)\equiv \half\theta_2^8(\tau),
\label{4.0}
\end{equation}
The reason for introducing 8th power of $\theta$ will be explained later,
 but as preparation of the discussion, we mention the 
structure of $q$-expansions of these functions.  
$\Theta_+(\tau) $,
$\Theta_2(\tau) $ and
$\Theta_-(\tau) $ have the following $q$-expansions
\begin{eqnarray}
&&\Theta_+(\tau)=1+q^n~{\rm terms},\no\\
&&\Theta_2(\tau)=2^7 q+q^n~{\rm terms},\no\\
&&\Theta_-(\tau)=2^4 q^\half+q^{n+\half}~{\rm terms}.
\end{eqnarray}
Note that $\Theta_-(\tau) $ is the only odd type.
Using these $\theta$ functions,
one can construct blow-up formulas
by multiplying $\frac{1}{\eta^{16}(\tau)} $,
\begin{eqnarray}
&&{\tilde Z}_+(\tau)\equiv \frac{\Theta_+(\tau)}{\eta^{16}(\tau)},\no\\
&&{\tilde Z}_-(\tau)\equiv \frac{\Theta_-(\tau)}{\eta^{16}(\tau)},\no\\
&&{\tilde Z}_2(\tau)\equiv \frac{\Theta_2(\tau)}{\eta^{16}(\tau)}.
\end{eqnarray}
These blow-up functions have the following modular properties
\[
\left(
  \begin{array}{c}
{\tilde Z}_+(\tau+1)\\
{\tilde Z}_-(\tau+1)\\
{\tilde Z}_2(\tau+1)
  \end{array}
\right)
=
e^{-\frac{4\pi i}{3}}
\left(
  \begin{array}{ccc}
1&0&0\\
0&-1&0\\
0&0&1
  \end{array}
\right)
\left(
  \begin{array}{c}
{\tilde Z}_+(\tau)\\
{\tilde Z}_-(\tau)\\
{\tilde Z}_2(\tau)
  \end{array}
\right),
\]

\begin{equation}
\left(
  \begin{array}{c}
{\tilde Z}_+(-\frac{1}{\tau})\\
{\tilde Z}_-(-\frac{1}{\tau})\\
{\tilde Z}_2(-\frac{1}{\tau})
  \end{array}
 \right)
=
\left(\sqrt{\frac{\tau}{i}} \right)^{-8}
\left(
  \begin{array}{ccc}
\half&\half&1\\
\half&\half&-1\\
\half&-\half&0
  \end{array}
\right)
\left(
  \begin{array}{c}
{\tilde Z}_+(\tau)\\
{\tilde Z}_-(\tau)\\
{\tilde Z}_2(\tau)
  \end{array}
\right).
\end{equation}
Thus, we can see that the above three functions form closed orbit under 
$S$-duality transformation. Note that if we require the length of the orbit 
to be three (in general, it becomes six), the only allowed power of $\theta$
is 8.
Especially, we have to note that there is an equality:
\begin{equation}
{\tilde Z}_2(-\frac{1}{\tau})
\left(
{\tilde Z}_+(-\frac{1}{\tau})-
{\tilde Z}_-(-\frac{1}{\tau})
\right)
=
\left(\sqrt{\frac{\tau}{i}} \right)^{-16}
{\tilde Z}_2(\tau)
\left(
{\tilde Z}_+(\tau)-
{\tilde Z}_-(\tau)
\right).
\label{kimo}
\end{equation}
\subsection{$K3$ Partition Function}
\label{sec:4-2}
In this subsection, we reconstruct the $K3$ partition functions
using the materials we have prepared \cite{vafa-witten}.

\paragraph{From $Z_{even}^{S_{0}}(\tau) $ and $Z_{odd}^{S_{0}}(\tau) $
to $Z_{even}^{K3}(\tau) $ and $Z_{odd}^{K3}(\tau) $}
~\\
In this part,
we derive $Z_{even}^{K3}(\tau) $ and $Z_{odd}^{K3}(\tau) $
from $Z_{even}^{S_{0}}(\tau) $ and $Z_{odd}^{S_{0}}(\tau) $ 
using ${\cal O}(-2)$ curve blow-up formula (\ref{4.0}).
To determine 
$Z_{even}^{K3}(\tau) $ and $Z_{odd}^{K3}(\tau) $
, we use the following two properties
\cite{vafa-witten}.
First property is that 
$Z_{even}^{K3}(\tau) $ and $Z_{odd}^{K3}(\tau) $
have $q$-expansions in the form of 
$(1+q^n {\rm ~terms})$ and of $(q^{-\half}+q^{\half+n} {\rm ~terms})$
respectively. This follows from dimensional counting of the moduli space. 
The next property is that
$Z_{even}^{K3}(\tau)-Z_{odd}^{K3}(\tau) $
is modular covariant under $\tau\to-\frac{1}{\tau} $ 
(see (\ref{zkm}) in Appendix B).
One can find
\begin{eqnarray}
Z_{odd}^{S_{0}}\frac{1}{\eta^{32}(\tau)}
&=&
q^{-\frac{1}{6}}q^{-\frac{32}{24}}+\cdots
\nonum
\\
&=&
q^{-\frac{3}{2}}+\cdots .
\label{4.1}
\end{eqnarray}

Thus to construct $Z_{even}^{K3}(\tau) $
from $Z_{odd}^{S_{0}}(\tau)$,
one blow-ups only by  ${\tilde Z}_2(\tau){\tilde Z}_-(\tau) $.
To construct $Z_{odd}^{K{3}}(\tau) $
from $Z_{odd}^{S_{0}}(\tau) $
one can blow-up  by either  
${\tilde Z}_2(\tau){\tilde Z}_+(\tau) $
or ${\tilde Z}_2(\tau){\tilde Z}_2(\tau) $
due to the first property.
Due to the second property
we choose the former. 
The same line of thought can be applied to the case of blowing up 
$Z_{even}^{S_{0}}(\tau) $ to $Z_{even}^{K{3}}(\tau) $
and to $Z_{odd}^{K{3}}(\tau) $.

Therefore, we are led to the following: 
\begin{eqnarray}
&&Z^{K{3}}_{even}(\tau)=
N\left(
Z^{S_{0}}_{odd}(\tau){\tilde Z}_2(\tau){\tilde Z}_-(\tau)
+
Z^{S_{0}}_{even}(\tau){\tilde Z}_2(\tau){\tilde Z}_+(\tau)
\right),\no\\
&&Z^{K3}_{odd}(\tau)=
N\left(
Z^{S_{0}}_{odd}(\tau){\tilde Z}_2(\tau){\tilde Z}_+(\tau)
+
Z^{S_{0}}_{even}(\tau){\tilde Z}_2(\tau){\tilde Z}_-(\tau)
\right).
\label{evev}
\end{eqnarray} 
Note that
$
Z^{S_{0}}_{even}(\tau)
-
Z^{S_{0}}_{odd}(\tau)
$
(\ref{zsm})
and
$ {\tilde Z}_2(\tau)  ({\tilde Z}_+(\tau)-{\tilde Z}_-(\tau))$(\ref{kimo})
are also modular covariant under $\tau\to-\frac{1}{\tau} $.

Here we remark two points concerning $q$-expansions.
The first point is that all materials of $Z^{K3}_* $ naturally have the form
either $q^n $ or $q^{n+\half}$,
since $Z^{K3}_* $ itself has the form either $q^n $ or $q^{n+\half}$.
This is why we only introduce $Z_{even}^{S_{0}} $,$Z_{odd}^{S_{0}}$ and 
$Z_{0}^{S_{0}} $
in \ref{sec:3-2} and ${\tilde Z}_\pm $ and ${\tilde Z}_2 $ in \ref{sec:4-1}.
The second point is that there should be one
which has the form beginning with $q $,
since the gap between $q^{-\frac{3}{2}} $ in (\ref{4.1}) 
and $1 $ in $Z^{K3}_{even} $ can only be filled with the combination 
$q $ and $q^\half $.
Therefore, we introduce 8-th power of $\theta_2 $ in (\ref{4.0}).

\paragraph{$Z^{K3}_0(\tau) $ and $Z^{K3}_{SO(3)}(\tau) $ }
~\\
Following the discussion in \cite{vafa-witten}
for $Z^{K3}_0(\tau)$, we set the form of $Z^{K3}_{even}(\tau)$
as follows:
\begin{equation}
Z^{K3}_0(\tau)=
\frac{N_1}{4}G(2\tau){\tilde Z}_a(\tau)
+N\left(
Z^{S_{0}}_{even}(\tau){\tilde Z}_2(\tau){\tilde Z}_-(\tau)
+
Z^{S_{0}}_{odd}(\tau){\tilde Z}_2(\tau){\tilde Z}_+(\tau)
\right),
\label{0}
\end{equation}
where ${\tilde Z}_a(\tau) $ is made of 
${\tilde Z}_\pm(\tau),{\tilde Z}_2(\tau)$
so that $Z^{K3}_0(\tau)$ is even type.

As the last step, we have to determine the unknown factors
$N,N_1 $ and the form of ${\tilde Z}_a(\tau) $.
Here
it is convenient to introduce
\begin{eqnarray}
&&{\tilde Z}_b(\tau)\equiv 
2{\tilde Z}_2(\tau)
\left(
{\tilde Z}_+(\tau)+{\tilde Z}_-(\tau)
\right)
=\frac{\theta_2^8(\tau)\theta_3^8(\tau)}{\eta^{32}(\tau)},\no\\
&&{\tilde Z}_c(\tau)\equiv
2{\tilde Z}_2(\tau)
\left(
{\tilde Z}_+(\tau)-{\tilde Z}_-(\tau)
\right)
=\frac{\theta_2^8(\tau)\theta_4^8(\tau)}{\eta^{32}(\tau)}.
\label{bc}
\end{eqnarray}
Using these functions, we have the following formulas:
\begin{eqnarray}
&&Z^{K3}_{even}(\tau)=
\frac{N}{4}
\left(
G(\frac{\tau}{2}){\tilde Z}_b(\tau)
-e^{\frac{\pi i}{3}}G(\frac{\tau}{2}+\frac{1}{2}){\tilde Z}_c(\tau)
\right),\no\\
&&Z^{K3}_{odd}(\tau)=
\frac{N}{4}
\left(
G(\frac{\tau}{2}){\tilde Z}_b(\tau)
+e^{\frac{\pi i}{3}}G(\frac{\tau}{2}+\frac{1}{2}){\tilde Z}_c(\tau)
\right),\no\\
&&Z^{K3}_0(\tau)=
\frac{N_1}{4} G(2\tau){\tilde Z}_a(\tau)
+\frac{N}{4}
\left(
G(\frac{\tau}{2}){\tilde Z}_b(\tau)
-e^{\frac{\pi i}{3}}G(\frac{\tau}{2}+\frac{1}{2}){\tilde Z}_c(\tau)
\right).
\label{eo0}
\end{eqnarray}
We also define $Z^K_{SO(3)}(\tau)$
\begin{eqnarray*}
Z^{K3}_{SO(3)}(\tau)
&=&
  Z^{K3}_0(\tau)+n_{even}Z^{K3}_{even}(\tau)+n_{odd}Z^{K3}_{odd}(\tau)
\\
&=&
\frac{N_1}{4} G(2\tau){\tilde Z}_a(\tau)
+
2^{20}N
G(\frac{\tau}{2}){\tilde Z}_b(\tau)
-2^9Ne^{\frac{\pi i}{3}}
G(\frac{\tau}{2}+\frac{1}{2}){\tilde Z}_c(\tau),
\end{eqnarray*}
where we use the numbers $n_0=1$,
$
n_{even}=\frac{2^{22}+2^{11}}{2}-1,
n_{odd}=\frac{2^{22}-2^{11}}{2}
$
(see Appendix A).

Following \cite{vafa-witten}, 
the ${\tilde Z}_a(\tau)$
should be transformed to ${\tilde Z}_b(\tau)$
under transformation $\tau\rightarrow -\frac{1}{\tau} $.
Therefore, we obtain the form,
\begin{equation}
{\tilde Z}_a(\tau)
=\frac{\theta_3^8(\tau)\theta_4^8(\tau)}{\eta^{32}(\tau)}.
\end{equation}
Then we apply the $S$-duality property (\ref{so3}),
\begin{equation}
Z^{K3}_0(-\frac{1}{\tau})=2^{-11}\tau^{-12}Z^K_{SO(3)}(\tau),
\end{equation}
that reduces to the equality:
\begin{eqnarray}
  Z^{K3}_0(-\frac{1}{\tau})=\tau^{-12}
\left(
\frac{N_1}{4}
2^4G(\frac{\tau}{2}){\tilde Z}_b(\tau)
+\frac{N}{4}2^{-4}G(2\tau){\tilde Z}_a(\tau)
-\frac{N}{4}e^{\frac{\pi i}{3}}G(\frac{\tau}{2}+\frac{1}{2})
{\tilde Z}_c(\tau)
\right).\no
\end{eqnarray}
Thus we get
\[
N_1=2^7N.
\]
Now we  compare the results with the ones in \cite{vafa-witten}.
We use the identities of theta and eta functions, 
\[
{\tilde Z}_a(\tau)=\frac{1}{\eta^{16}(2\tau)},
\]
\[
{\tilde Z}_b(\tau)=\frac{2^8}{\eta^{16}(\frac{\tau}{2})},
\]
\[
{\tilde Z}_c(\tau)=e^{\frac{8\pi i}{3}}
\frac{2^8}{\eta^{16}(\frac{\tau}{2}+\half)},
\]
and rewrite $Z^{K3}_0,Z^{K3}_{even},Z^K_{odd} $ and $Z^{K3}_{SO(3)} $.
After all, our results turn into, 
\begin{eqnarray}
&&Z^{K3}_0(\tau)=2^7N\left(
\frac{1}{4}\frac{1}{\eta^{24}(2\tau)}
+\half\frac{1}{\eta^{24}(\frac{\tau}{2})}
+\half\frac{1}{\eta^{24}(\frac{\tau}{2}+\half)}
\right),\no\\
&&Z^{K3}_{even}(\tau)=2^7N\left(
\half\frac{1}{\eta^{24}(\frac{\tau}{2})}
+\half\frac{1}{\eta^{24}(\frac{\tau}{2}+\half)}
\right),\no\\
&&
Z^{K3}_{odd}(\tau)=2^7N\left(
\half\frac{1}{\eta^{24}(\frac{\tau}{2})}
-\half\frac{1}{\eta^{24}(\frac{\tau}{2}+\half)}
\right),\no\\
&&Z^{K3}_{SO(3)}(\tau)=2^7N\left(
\frac{1}{4}\frac{1}{\eta^{24}(2\tau)}
+2^{21}\frac{1}{\eta^{24}(\frac{\tau}{2})}
+2^{10}\frac{1}{\eta^{24}(\frac{\tau}{2}+\half)}
\right).
\end{eqnarray}
If we set $2^7 N=1 $, the results are
the same as Vafa-Witten's \cite{vafa-witten}.
Thus we have connected $T^4$ with $K3$
by orbifold construction at the partition function level.

\section{Conclusion}
In this paper, we reconstruct the partition function of twisted $N=4$ SYM 
theory on $K3$ surface using the orbifold construction $T^{4}/{\bf Z}_{2}$ 
at the partition function level.
But, we used two major assumptions:\\
(i) The contribution from the untwisted sector is fundamentally described by 
G\"ottsche formula applied naively to even part of $H^{*}(T^{4},{\bf C})$.\\
(ii) ${\cal O}(-2)$ curve blow-up formula is given by 
${\theta_{i}(\tau)}/{(\eta(\tau))^{2}}$. \\
Justification of these two things remains to be pursued. Roughly speaking, 
we have neglected the behavior of vector bundles on sixteen singularity 
points of $S_{0}$. Since singularity points have complex codimension $2$ in base 
manifold, we expected that they do not affect the theory severely unlike 
cosmic strings.  
   
Now, we discuss the future direction of our computations. 
In this paper, we were only considering the partition function 
for $SU(2)$ or $SO(3)$,
but there are methods to obtain the partition functions
for the other gauge groups, such as $SU(N) $ or $SU(N)/{\bf Z}_N$
\cite{vafa-witten}, \cite{lozano}, \cite{bonelli}.
Our computations can be also generalized to the other gauge groups.
One more possible 
generalization is changing ${\bf Z}_{2}$ into other discrete groups: 
for example, A-D-E type discrete subgroup of $SU(2)$. One aim of this 
generalization is to construct various blow-up formulas for A-D-E 
singularities. We think that this direction is interesting with respect to 
the connection with the work of Nakajima on ALE spaces. We also hope our 
approach will contribute to the production of many examples of $S$-duality 
conjecture of twisted $N=4$ SYM theory. 
 
\label{sec:4}

{\bf Acknowledgment}\\
We would like to give many thanks to N. Kawamoto, K. Ono, T. Eguchi and 
K. Yoshioka for comments and discussions. 

\appendix
\renewcommand{\theequation}{\alph{section}.\arabic{equation}}

\section{Counting $n_0$, $n_{even}$ and $n_{odd}$}
\label{sec:a}
\setcounter{equation}{0}

In this section, we derive the numbers 
$n_0$,$n_{even}$ and $n_{odd}$ of $K3$,
 following the discussion in \cite{vafa-witten}.
We denote $v\in H^2(K3,Z_2)$ by 
\begin{equation}
v=\sum_{i=1}^{22}\epsilon_i v_i,
\end{equation}
where $v_i\in H^2(X,Z) ~i=1,\ldots ,22$
and $\epsilon_i=0~{\rm or}~1. $
Thus $v$ can take $2^{22} $ values. 

For $u,v\in H^2(K3,Z_2) $, intersection form $u\cdot v$
on $K3 $ is $H^{\oplus 3} \oplus (-E_8)^{\oplus 2}$, but
for our purpose to count the number of $0,$ even and odd types,
it is sufficient to think that intersection form is $H^{\oplus 11}$, 
due to the discussion in \cite{vafa-witten}.
Thus we separate $v_i\in H^2(K3,Z) $ as
\begin{equation}
(v_{2j-1},v_{2j}),~j=1,\ldots ,11,
\end{equation}
and we set that intersection form of each piece
is $H$, that is
\begin{equation}
\left(
\begin{array}{cc}
v_{2j-1}\cdot v_{2j-1}& v_{2j-1}\cdot v_{2j}\\
v_{2j}\cdot v_{2j-1}& v_{2j}\cdot v_{2j}
\end{array}
\right)
=
\left(
\begin{array}{cc}
0&1\\
1&0
\end{array}
\right).
\end{equation}             
For each piece $v_{2j-1},v_{2j}$, we define
\begin{equation}
V_j=\epsilon_{2j-1}v_{2j-1}+\epsilon_{2j}v_{2j}.
\end{equation}
If $V_j^2=0~{\rm mod}~4$, then $V_j =0,v_{2j-1}~{\rm or}~v_{2j}$. 
If $V_j^2=2~{\rm mod}~4$, then $V_j =v_{2j-1}+v_{2j}$. 
Using this, we count the numbers 
$n_0$, $n_{even}$ and $n_{odd}$ of $K3$.

\begin{eqnarray}
n_0
&=&
\mbox{\symbol{35} of}~v=0
\nonum
\\
&=& 1,
\\
& &\nonum
\\
n_{even}
&=&
\mbox{\symbol{35} of}~v^2=0~{\rm mod}~4~{\rm but~nonzero}
\nonum
\\
&=& (
\mbox{\symbol{35} of}~{\rm even~choices~of}~v_{2j-1}+v_{2j}~
{\rm in}~H^{\oplus 11} )-1
\nonum
\\
&=&
{}_{11}C_03^{11}+\cdots +{}_{11}C_{10}3^1-1
\nonum
\\
&=&
\half(2^{22}+2^{11})-1,
\\
& &\nonum
\\
n_{odd}
&=&
\mbox{\symbol{35} of}~v^2=2~{\rm mod}~4
\nonum
\\
&=& 
\mbox{\symbol{35} of}~{\rm odd~choices~of}~v_{2j-1}+v_{2j}~
{\rm in}~H^{\oplus 11} 
\nonum
\\
&=&
{}_{11}C_13^{10}+\cdots +{}_{11}C_{11}3^0
\nonum
\\
&=&
\half(2^{22}-2^{11}).
\end{eqnarray}

\section{Modular Properties of $Z_v^{K3}(\tau) $}
\label{sec:b}
\setcounter{equation}{0}

In this section, we derive 
the useful modular properties of $Z_v^{K3}(\tau) $
on $K3$ from (\ref{vw}).

\begin{equation}
Z^{K3}_v\left(-\frac{1}{\tau}\right)=2^{-11}
\left(
\frac{\tau}{i}
\right)^{-12}
\cdot
\sum_{u\in H^2(K3,Z_2)}
(-1)^{u\cdot v}Z^{K3}_u(\tau).
\end{equation}

From the definition of 
$Z^{K3}_{SO(3)}(\tau)\equiv \sum_{u\in H^2(K3,Z_2)} Z_u^{K3}(\tau)$, 
it is apparent 
\begin{equation}
Z^{K3}_0\left(-\frac{1}{\tau}\right)=2^{-11}
\left(
\frac{\tau}{i}
\right)^{-12}
Z^{K3}_{SO(3)}(\tau).
\end{equation}

For the modular properties of even or odd type,
it is necessary to count the numbers 
$n_{even}^{u\cdot v=1},n_{even}^{u\cdot v=0}, 
n_{odd}^{u\cdot v=1} $ and $n_{odd}^{u\cdot v=0} $
defined below. For even $v$, $Z^{K3}_{odd}(\tau) $ 
transform

\begin{eqnarray}
& &
Z^{K3}_{odd}\left(-\frac{1}{\tau}\right)
\nonum
\\
&=&
2^{-11}
\left(
\frac{\tau}{i}
\right)^{-12}
\cdot\left(
Z^{K3}_0(\tau)
+\sum_{u:{\rm even}}
(-1)^{u\cdot v}Z^{K3}_{even}(\tau)
+\sum_{u:{\rm odd}}
(-1)^{u\cdot v}Z^{K3}_{odd}(\tau)
\right)
\nonum
\\
&=&
2^{-11}
\left(
\frac{\tau}{i}
\right)^{-12}
\cdot\left(
Z^{K3}_0(\tau)
+(-n_{even}^{u\cdot v=1}+n_{even}^{u\cdot v=0})
Z^{K3}_{even}(\tau)
+(-n_{odd}^{u\cdot v=1}+n_{odd}^{u\cdot v=0})
Z^{K3}_{odd}(\tau)
\right),
\nonum
\\
\end{eqnarray}
where we separate summation $\sum_{u:{\rm even}}$
to $u\cdot v=1~ {\rm mod}~2 $ type
and  
$u\cdot v=0~ {\rm mod}~2 $ type
and introduce the numbers
$n_{even}^{u\cdot v=1}$ and $n_{even}^{u\cdot v=0}$.
The numbers
$n_{odd}^{u\cdot v=1}$ and $n_{odd}^{u\cdot v=0}$
are defined from summation $\sum_{u:{\rm odd}}$ similarly.
We count the numbers 
$n_{even}^{u\cdot v=1},n_{even}^{u\cdot v=0}, 
n_{odd}^{u\cdot v=1} $ and $n_{odd}^{u\cdot v=0} $ on $K3$.
Without loss of generality, we take odd $v(=v_1+v_2 )$.

\begin{eqnarray}
  n_{even}^{u\cdot v=1}
&=&
\mbox{\symbol{35} of}~u\cdot v=1~{\rm mod}~2~ {\rm in~even~}u\mbox{'s} 
\nonum
\\
&=&
\mbox{\symbol{35} of}~u=v_1~{\rm or}~v_2+{\rm even }~u~{\rm combinations}
\nonum
\\
&=&
2({}_{10}C_03^{10}+\cdots +{}_{10}C_{10}3^0)
\\
&=&
2^{20}+2^{10},\no\\
&&\no\\ 
n_{even}^{u\cdot v=0}
&=&
\mbox{\symbol{35} of}~u\cdot v=0~{\rm mod}~2~ {\rm in~even~}u\mbox{'s}
\nonum
\\
&=&
(
\mbox{\symbol{35} of}~u=0+{\rm even }~u~{\rm combinations~nonzero}
)
\nonum
\\
&&+
(
\mbox{\symbol{35} of}~u=v_1+v_2+{\rm even }~u~{\rm combinations}
)
\nonum
\\
&=&
({}_{10}C_03^{10}+\cdots +{}_{10}C_{10}3^0)-1
+({}_{10}C_13^9+\cdots +{}_{10}C_93^1)
\nonum
\\
&=&
2^{20}-1,\\
&&\no\\
 n_{odd}^{u\cdot v=1}
&=&
\mbox{\symbol{35} of}~u\cdot v=1~{\rm mod}~2~ {\rm in~odd~}u \mbox{'s}
\nonum
\\
&=&
\mbox{\symbol{35} of}~u=v_1~{\rm or}~v_2+{\rm odd }~u~{\rm combinations}
\nonum
\\
&=&
2({}_{10}C_13^9+\cdots +{}_{10}C_93^1 ) 
\nonum
\\
&=&
2^{20}-2^{10},\\
&&\no\\
  n_{odd}^{u\cdot v=0}
&=&
\mbox{\symbol{35} of}~u\cdot v=0~{\rm mod}~2~ {\rm in~odd~}u \mbox{'s}
\nonum
\\
&=&
(
\mbox{\symbol{35} of}~u=0+{\rm odd }~u~{\rm combinations}
)
\nonum
\\
&&+
(
\mbox{\symbol{35} of}~u=v_1+v_2+{\rm odd }~u~{\rm combinations}
)
\nonum
\\
&=&
({}_{10}C_03^{10}+\cdots +{}_{10}C_{10}3^0)
+({}_{10}C_13^9+\cdots +{}_{10}C_{9}3^1)
\nonum
\\
&=&
2^{20}.
\end{eqnarray}

Finally we obtain the modular property of $Z^{K3}_{odd}(\tau) $

\begin{eqnarray}
& &
Z^{K3}_{odd}\left(-\frac{1}{\tau}\right)
\nonum
\\
&=&
2^{-11}
\left(
\frac{\tau}{i}
\right)^{-12}
\cdot\left(
Z^{K3}_0(\tau)
+(-2^{10}-1)
Z^{K3}_{even}(\tau)
+2^{10}
Z^{K3}_{odd}(\tau)
\right).
\end{eqnarray}

In the same way as $Z^{K3}_{odd}(\tau)$, we obtain
the modular property of $Z^{K3}_{even}(\tau) $

\begin{eqnarray}
& &
Z^{K3}_{even}\left(-\frac{1}{\tau}\right)
\nonum
\\
&=&
2^{-11}
\left(
\frac{\tau}{i}
\right)^{-12}
\cdot\left(
Z^{K3}_0(\tau)
+(2^{10}-1)
Z^{K3}_{even}(\tau)
-2^{10}
Z^{K3}_{odd}(\tau)
\right).
\end{eqnarray}

Combining $Z^{K3}_{odd}(\tau)$ and $Z^{K3}_{even}(\tau) $, we also obtain

\begin{eqnarray}
Z^{K3}_{even}\left(-\frac{1}{\tau}\right)
-
Z^{K3}_{odd}\left(-\frac{1}{\tau}\right)
&=&
\left(
\frac{\tau}{i}
\right)^{-12}
\cdot\left(
Z^{K3}_{even}(\tau)
-
Z^{K3}_{odd}(\tau)
\right).
\label{zkm}
\end{eqnarray}

\end{document}